\documentclass[reqno, centertags, 11pt]{amsart}
\usepackage{amsmath,amsthm,amsfonts,amssymb}
\usepackage{graphicx}
\usepackage{epstopdf}
\usepackage{comment}

\renewcommand{\part}[1]{\left(#1\right)}  
\newcommand{\parg}[1]{\left\{#1\right\}}  
\newcommand{\ket}[1]{\vert {#1} \rangle}  

\newcommand{\nl}{\par\noindent}   
\newcommand{\Prob}{{\tt p}}    

\newcommand{\SId}{\sqrt{ \rm I}}

\newcommand{\C}{\mathbb{C}}

\newcommand{\Id}{{\rm I}}

\newtheorem{theorem}{Theorem}[section]

\newtheorem{definition}{Definition}[section]
\newtheorem{example}{Example}[section]

\numberwithin{equation}{section}

{\begin{proof}[#1]%
\renewcommand{\qed}{\rule{0mm}{1mm}\hfill{\footnotesize\textsc{q.e.d.}\
(#2)}}}
{\end{proof}
}

\begin{document}
   \title[Epistemic semantics. Part I]
 {\textbf{A quantum computational semantics for epistemic logical operators.
 \nl Part I: Epistemic structures}}

\author[Beltrametti]{Enrico Beltrametti}
\address[Beltrametti]{Dipartimento di Fisica,
Universit\`a di Genova,
  Via Dodecaneso, 33,
I-16146 Genova, Italy} \email{enrico.beltrametti@ge.infn.it}

\author[Dalla Chiara]{Maria Luisa Dalla Chiara}
\address[Dalla Chiara]{Dipartimento di Filosofia,
   Universit\`a di Firenze,
   Via Bolognese 52, I-50139 Firenze, Italy}
\email{dallachiara@unifi.it}

\author[Giuntini]{Roberto Giuntini}
\address[R.~Giuntini]{Dipartimento di Filosofia,
    Universit\`a di Cagliari,
    Via Is Mirrionis 1, I-09123 Cagliari, Italy.}
    \email{giuntini@unica.it}


\author[Leporini]{Roberto Leporini}
\address[R. Leporini]{Dipartimento di Ingegneria,
    Universit\`a di Bergamo,
    viale Marconi 5, I-24044 Dalmine (BG), Italy.}
\email{roberto.leporini@unibg.it}

\author[Sergioli]{Giuseppe Sergioli}
\address[G.~Sergioli]{Dipartimento di Filosofia,
    Universit\`a di Cagliari,
    Via Is Mirrionis 1, I-09123 Cagliari, Italy.}
    \email{giuseppe.sergioli@gmail.com}

\maketitle

\begin{abstract}
Some critical open problems of epistemic logics can be
investigated in the framework of a quantum computational approach.
The basic idea is to interpret sentences like ``Alice knows that
Bob does not understand that $\pi$ is irrational'' as pieces of
quantum information (generally represented by density operators of
convenient Hilbert spaces). Logical epistemic operators ({\em to
understand\/}, {\em to know\/} ...) are dealt with as (generally
irreversible) quantum operations, which are, in a sense, similar
to measurement-procedures. This approach permits us to model some
characteristic epistemic processes, that concern both human and
artificial intelligence.  For instance, the operation of
``memorizing and retrieving information'' can be formally
represented, in this framework, by using a quantum teleportation
phenomenon.
\end{abstract}

\section{Introduction}
Logical theories of epistemic operators ({\em to know\/}, {\em to
believe\/},...) have given rise to a number of interesting open
questions. Most standard approaches (based on extensions of
classical logic) succeed in modelling a general notion of
``potential knowledge''. In this framework, a sentence like
``Alice knows that $\pi$ is irrational'' turns out to have the
meaning ``Alice {\em could know\/} that $\pi$ is irrational'',
rather than ``Alice {\em actually knows\/} that $\pi$ is
irrational''. A consequence of such a strong characterization of
knowledge  is the unrealistic phenomenon of {\em logical
omniscience\/}, according to which knowing a sentence  implies
knowing all its logical consequences.

A weaker approach to epistemic logics can be developed in the
framework of a {\em quantum computational semantics\/}. The aim is
trying  to describe forms of ``actual knowledge'', which should
somehow reflect the real limitations both of human and of
artificial intelligence.

In quantum computational semantics meanings of sentences are
represented as pieces of quantum information (mathematically
described as  density operators living in convenient Hilbert
spaces), while  the logical connectives correspond to special
examples of quantum logical gates. How to interpret, in this
framework, epistemic sentences like ``Alice knows that Bob does
not understand  that $\pi$ is irrational''? The leading idea can
be sketched as follows. The semantics is based on abstract
structures that contain finite sets of {\em epistemic agents\/}
evolving in time. Each agent (say, Alice at a particular time) is
characterized by two fundamental epistemic parameters:
\begin{itemize}
\item a set of density operators, representing the information
    that is accessible to our agent.
    \item A ``truth-conception'' (called the {\em
        truth-perspective\/} of the agent in question), which
        is technically determined by the choice of an
        orthonormal basis of the two-dimensional Hilbert space
        $\C^2$. In this way, any pair of qubits, corresponding
        to the elements of the basis that has been chosen, can
        be regarded  as a particular idea about the
        truth-values {\em Truth\/} and {\em Falsity\/}. From a
        physical point of view, we can imagine that a
        truth-perspective is
  associated to a physical apparatus that permits one to
measure a given observable.

\end{itemize}

The knowledge operations, described in this semantics, turn out to
be deeply different from  quantum logical gates, since  they
cannot be, generally, represented by unitary quantum operations.
The ``act of knowing'' seems to involve some intrinsic
irreversibility, which is, in a sense, similar to what happens in
the case of measurement-procedures.

The first Part of this article is devoted to a mathematical
description of the notion of {\em epistemic structure\/} in a
Hilbert-space environment, while  the semantics for an epistemic
quantum computational language is developed in the second Part. We
will analyze, in this framework,  some  epistemic situations that
seem to characterize  ``real'' processes of acquiring and
transmitting information.

\section{Quantum information and truth-perspectives}
We will first recall some basic notions of quantum computation
that  will be used in our  semantics. The general mathematical
environment is the $n$-fold tensor product of the Hilbert space
$\C^2$:
$$\mathcal H^{(n)}:=
\underbrace{\C^2\otimes\ldots\otimes\C^2}_{n-times},$$ where all
pieces of quantum information live. The elements $\ket{1} = (0,1)$
and $\ket{0} = (1,0)$ of the canonical orthonormal basis $B^{(1)}$
of $\C^2$ represent, in this framework, the two classical bits,
which can be also regarded as the canonical truth-values {\em
Truth\/} and {\em Falsity\/}, respectively.
  The canonical basis of $\mathcal H^{(n)}$ is the set $$B^{(n)} =
\parg{\ket{x_1}\otimes \ldots \otimes \ket{x_n}: \ket{x_1},
\ldots, \ket{x_n} \in B^{(1)}}.$$ As usual, we will briefly write
$\ket{x_1, \ldots, x_n}$ instead of $\ket{x_1} \otimes \ldots
\otimes \ket{x_n}$. By definition, a {\em quregister\/} is a unit
vector of $\mathcal H^{(n)}$. Quregisters thus correspond to pure
states, namely to maximal pieces of information about the quantum
systems that are supposed to store a given amount of quantum
information. We shall also make reference to {\em mixtures\/} of
quregisters, to be called {\em qumixes\/}, associated to density
operators $\rho$ of $\mathcal H^{(n)}$. We will denote by
$\mathfrak D(\mathcal H^{(n)})$ the set of all qumixes of
$\mathcal H^{(n)}$, while $\mathfrak D= \bigcup_n\parg{\mathfrak
D(\mathcal H^{(n)})}$ will represent the set of all possible
qumixes. Of course, quregisters can be represented as special
cases of qumixes, having the form $P_{\ket{\psi}}$ (the projection
over the one-dimensional closed subspace determined by the
quregister $\ket{\psi}$).

From an intuitive point of view, a basis-change in $\C^2$ can be
regarded as a change of our {\em truth-perspective\/}. While in
the canonical case, the truth-values {\em Truth\/} and {\em
Falsity\/} are identified with the two classical bits $\ket{1}$
and $\ket{0}$, assuming a different basis corresponds to a
different idea of {\em Truth\/} and {\em Falsity\/}. Since any
basis-change in $\C^2$ is determined by a unitary operator, we can
identify a {\em truth-perspective\/} with
 a unitary operator $\frak T$ of $\C^2$.
 We will write:
$$ \ket{1_{\frak T}}= \frak T \ket{1};\;  \ket{0_{\frak T}}=
\frak  T \ket{0},$$ and we will assume that $ \ket{1_{\frak T}}$  and
$\ket{0_{\frak T}}$ represent respectively the truth-values {\em
Truth\/}  and {\em Falsity\/} of the truth-perspective $\frak T$.
The {\em canonical truth-perspective\/} is, of course, determined
by the identity operator $\Id$ of $\C^2$. We will indicate by
$B^{(1)}_\frak T$ the orthonormal basis determined by $\frak T$;
while $B^{(1)}_\Id$ will represent the canonical basis.

Any unitary operator $\frak T$ of $\mathcal H^{(1)}$ can be
naturally extended to a unitary operator $\frak T^{(n)}$ of
$\mathcal H^{(n)}$ (for any $n \geq 1$):

$$\frak T^{(n)}\ket{x_1,\ldots,x_n}= \frak T\ket{x_1}\otimes \ldots
\otimes \frak T\ket{x_n}.$$

Accordingly, any choice of a unitary operator $\frak T$ of
$\mathcal H^{(1)}$
 determines an orthonormal
basis  $B^{(n)}_\frak T$ for $\mathcal H^{(n)}$ such that:

$$B^{(n)}_\frak T =
\parg{\frak T^{(n)}\ket{x_1,\ldots,x_n}:\ket{x_1,\ldots,x_n} \in B_\Id^{(n)}
}.$$ Instead of  $\frak T^{(n)}\ket{x_1,\ldots,x_n}$ we will also write
$\ket{x_{1_{\frak T}},\ldots,x_{n_{\frak T}}}$.

 The elements of $B^{(1)}_\frak T$ will be called the $\frak T$-{\it bits}
    of $\mathcal H^{(1)}$;  while the elements of
    $B^{(n)}_\frak T$ will represent the $\frak T$-{\it registers} of
    $\mathcal H^{(n)}$.

On this ground the notions of {\em truth\/}, {\em falsity\/} and
{\em probability\/} with respect to any  truth-perspective
$\mathfrak T $ can be defined in a natural way.

 \begin{definition} {\em ($\frak T$-true  and
$\frak T$-false registers)}\label{def:true} \nl
\begin{itemize}
\item $\ket{x_{1_{\frak T}},\ldots,x_{n_{\frak T}}}$ is a {\em
    $\frak T$-true register} iff $\ket{x_{n_{\frak T}}} =
    \ket{1_{\frak T}};$
\item $\ket{x_{1_{\frak T}},\ldots,x_{n_{\frak T}}}$ is a {\em
    $\frak T$-false register} iff $\ket{x_{n_{\frak T}}} =
    \ket{0_{\frak T}}.$

     \end{itemize}

           \end{definition}

In other words, the {\em $\frak T$-truth-value\/} of a $\frak
T$-register
 (which corresponds to a sequence of $\frak T$-bits) is determined by its
last element.\footnote{As we will see, the application of a
classical reversible gate to a register $\ket{x_1,\ldots,x_n}$
transforms the (canonical) bit $\ket{x_n}$ into the target-bit
$\ket{x_n^\prime}$, which behaves as the final truth-value. This
justifies our choice in Definition \ref{def:true}.}

\begin{definition} {\em ($\mathfrak T$-truth and $\mathfrak T$-falsity)}
 \nl
\begin{itemize}
\item The $\mathfrak T$-{\em truth} of $\mathcal H^{(n)}$   is
    the projection operator $^\mathfrak TP_1^{(n)}$ that
    projects over the closed subspace spanned by the set of
    all $\mathfrak T$- true registers;
\item the $\mathfrak T$-{\em falsity} of $\mathcal H^{(n)}$ is
    the projection operator  $^\mathfrak TP_0^{(n)}$ that
    projects over the closed subspace spanned by the set of
    all $\mathfrak T$- false registers.
    \end{itemize}
 \end{definition}

In this way, truth and falsity are dealt with as mathematical
representatives of possible physical properties. Accordingly, by
applying the Born-rule, one can naturally define the
probability-value of any qumix with respect to the
truth-perspective $\mathfrak T$.
\begin{definition} {\em (}$\mathfrak T$-{\em Probability)} \nl
For any $\rho \in \mathfrak D(\mathcal H^{(n)})$,
$${\tt p}_\mathfrak T(\rho) := {\tt Tr}(^\mathfrak TP_1^{(n)} \rho),$$
where ${\tt Tr}$ is the trace-functional.
\end{definition}
We interpret ${\tt p}_\mathfrak T(\rho)$ as the probability that
the information $\rho$ satisfies the $\mathfrak T$-Truth.

In the particular case of qubits, we will
 obviously  obtain:
$${\tt p}_\mathfrak T(a_0\ket{0_\mathfrak T} + a_1\ket{1_\mathfrak
T})
 =|a_1|^2.   $$

For any choice of a truth-perspective  $\mathfrak T$, the set
$\mathfrak D$ of all qumixes can be pre-ordered by a relation that
is defined in  terms of the probability-function ${\tt
p}_\mathfrak T$.

\begin{definition} {\em (Preorder)} \label{de:preordine}\nl
$\rho \preceq_\mathfrak T \sigma$ iff ${\tt p}_\mathfrak T(\rho)
\le {\tt p}_\mathfrak T(\sigma)$.
\end{definition}

As is well known, quantum information is processed by {\em quantum
logical gates\/} (briefly, {\em gates\/}): unitary operators that
transform quregisters into quregisters in a reversible way.  Let
us recall the definition of  some gates that play a special role
both from the computational and from the logical point of view.

\begin{definition}\label{de:not}{\em (The negation)}
\nl For any $n\geq 1$, the {\em negation} on $\mathcal H^{(n)}$ is
the linear operator ${\tt NOT}^{(n)}$ such that, for every element
$\ket{x_1,\ldots,x_n}$ of the  canonical basis,
$$
{\tt NOT}^{(n)} \ket{x_1,\ldots,x_n} = \ket{x_1,\ldots, x_{n-1}}\otimes
\ket{1-x_{n}}.
$$
\end{definition}

In particular, we obtain:
$${\tt NOT}^{(1)}\ket{0}= \ket{1}; \,\,
{\tt NOT}^{(1)}\ket{1}= \ket{0}, $$ according to the classical
truth-table of negation.

\begin{definition} {\em (The Toffoli gate)}\label{de:toffoli}\nl
For any $n,m,p\geq 1$, the {\em Toffoli gate\/} is the linear
operator ${\tt T}^{(n,m,p)}$ defined on $\mathcal H^{(n+m+p)}$
such that, for every element $\ket{x_1,\ldots, x_n}\otimes
\ket{y_1,\ldots, y_m}\otimes
       \ket{z_1,\ldots, z_p} $
of the  canonical basis,
\begin{multline*}
{\tt T}^{(n,m,p)} \ket{x_1,\ldots, x_n, y_1,\ldots,
y_m, z_1,\ldots, z_p}\\
=\ket{x_1,\ldots, x_n, y_1,\ldots, y_m,z_1,\ldots,
z_{p-1}}\otimes\ket{x_n y_m \widehat{+} z_p},
\end{multline*}
where $\widehat{+}$ represents the addition modulo $2$.
\end{definition}


 \begin{definition} {\em (The ${\tt XOR}$-gate)} \label{de:xor}\nl
For any $n,m\geq 1$, the {\em Toffoli gate\/} is the linear
operator ${\tt XOR}^{(n,m)}$ defined on $\mathcal H^{(n+m)}$ such
that, for every element $\ket{x_1,\ldots, x_n}\otimes
\ket{y_1,\ldots, y_m}$ of the  canonical basis,
$${\tt XOR}^{(n,m)} \ket{x_1,\ldots, x_n, y_1,\ldots,y_m}
=\ket{x_1,\ldots, x_n, y_1,\ldots, y_{m-1}}\otimes\ket{x_n \widehat{+} y_m},$$
where $\widehat{+}$ represents the addition modulo $2$.
 \end{definition}

 \begin{definition} {\em (The ${\tt SWAP}$-gate)} \label{de:swap} \nl
For any $n\geq 1$, for any $i$ and for any $j$ (where $1 \le i \le
n$ and $1 \le j \le n$), the {\em SWAP gate\/} is the linear
operator ${\tt SWAP}^{(n)}_{(i,j)}$ defined on $\mathcal H^{(n)}$
such that, for every element $\ket{x_1,\ldots,x_i,\ldots,
x_j,\ldots x_n}$ of the canonical basis,
$${\tt SWAP}^{(n)}_{(i,j)}
 \ket{x_1,\ldots,x_i, \ldots, x_j,\ldots, x_n}
=\ket{x_1,\ldots,x_j, \ldots, x_i,\ldots, x_n}.$$
In other words, ${\tt SWAP}^{(n)}_{(i,j)}$ exchanges the $i$-th
with the $j$-th element  in any element of the basis.
 \end{definition}

\begin{definition}{\em (The Hadamard-gate)}
\nl For any $n\geq 1$, the {\em Hadamard-gate} on $\mathcal
H^{(n)}$ is the linear operator $\SId^{(n)}$ such that for every
element $\ket{x_1,\ldots,x_n}$ of the  canonical basis:
\begin{equation*}
\SId^{(n)}\ket{x_1,\ldots,x_n}=
 \ket{x_1,\ldots,x_{n-1}}\otimes
  \frac{1}{\sqrt{2}}\left((-1)^{x_{n}}\ket{x_n} +\ket{1-x_{n}}\right).
\end{equation*}

  \end{definition}

In particular we obtain:
$$\SId^{(1)} \ket{0}= \frac{1}{\sqrt{2}}(\ket{0} + \ket{1});
\SId^{(1)} \ket{1}= \frac{1}{\sqrt{2}}(\ket{0} - \ket{1}). $$
Hence, $\SId^{(1)}$ transforms bits into genuine qubits.

\begin{definition}{\em (The square root of {\tt NOT})}\nl
For any $n\geq 1$, the {\em square root of {\tt NOT}} on $\mathcal
H^{(n)}$ is the linear operator $\sqrt{\tt NOT}^{(n)}$ such that
for every element $\ket{x_1,\ldots,x_n}$ of the canonical basis:
\begin{equation*}
\sqrt{\tt NOT}^{(n)}\ket{x_1,\ldots,x_n}=
\ket{x_1,\ldots,x_{n-1}}\otimes \left(\frac{1-i}{2}\ket{x_n} +\frac{1+i}{2}\ket{1-x_{n}}\right),
\end{equation*}
where $i=\sqrt{-1}$.
\end{definition}

All gates can be naturally transposed from the canonical
truth-perspective to any truth-perspective  $\mathfrak T$. Let
$G^{(n)}$ be any gate defined with respect to the canonical
truth-perspective. The {\em twin-gate\/} $G^{(n)}_{\mathfrak T}$,
defined with respect to the truth-perspective $\mathfrak T$, is
determined as follows:
$$
G^{(n)}_{\mathfrak T}:=
 \mathfrak T^{(n)} G^{(n)}\mathfrak T^{{(n)}^\dagger},   $$
 where
$\mathfrak T^{{(n)}^\dagger}$ is the adjoint of $\mathfrak T$.

All $\mathfrak T$-gates can be canonically extended to the set
$\mathfrak D$ of all qumixes.  Let $G_\mathfrak T$ be any gate
defined on $\mathcal H^{(n)}$. The corresponding {\em qumix
gate\/} (also called {\em unitary quantum operation\/})
$^\mathfrak DG_\mathfrak T$ is defined as follows for any $\rho
\in \mathfrak D(\mathcal H^{(n)})$:
$$^{\mathfrak D}G_\mathfrak T\rho=
 G_\mathfrak T\,\rho\, G_\mathfrak T^\dagger.  $$
It is interesting to consider a convenient notion of {\em
distance\/} between
  truth-perspectives.  As is well known, different definitions of distance
 between vectors can be found in the literature. For our aims it
 is convenient to adopt the Fubini-Study definition of distance between
  two qubits.
 \begin{definition} ({\em The Fubini-Study distance})
 \label{fubini} \nl
 Let $\ket{\psi}$ and $\ket{\varphi}$ be two qubits.
 $$d(\ket{\psi}, \ket{\varphi})=
 \frac{2}{\pi}\arccos|\langle \psi|\varphi\rangle|.  $$

\end{definition}

 This notion of distance satisfies the following conditions:
 \begin{enumerate}
 \item  $d(\ket{\psi}, \ket{\varphi})$ is a metric distance;
 \item $\ket{\psi} \perp  \ket{\varphi} \Rightarrow
     d(\ket{\psi}, \ket{\varphi})=1 $;
 \item $d(\ket{1}, \ket{1_{Bell}}) = \frac {1}{2}$, where
     $\ket{1}$  is the canonical truth, while $\ket{1_{Bell}}
     = \sqrt\Id^{(1)}\ket{1} = \left(\frac{1}{\sqrt{2}}, -
     \frac{1}{\sqrt{2}}  \right) $ represents the {\em
     Bell-truth\/} (which corresponds  to a maximal
 uncertainty with respect to the canonical truth).

 \end{enumerate}

  On this ground, one can naturally define the {\em epistemic
 distance\/}  between two  truth-perspectives.

 \begin{definition} ({\em Epistemic distance})
 \label{de:epdist}\nl
 Let $\mathfrak T_1$ and $\mathfrak T_2$  be two
 truth-perspectives. $$d^{Ep}(\mathfrak T_1, \mathfrak T_2)=
 d(\ket{1_{\mathfrak T_1}}, \ket{1_{\mathfrak T_2}}).   $$

 \end{definition}

 In other words, the epistemic distance  between the
 truth-perspectives $\mathfrak T_1$   and $\mathfrak T_2$
 is  identified with the distance between the two qubits
 that represent the truth-value {\em Truth\/} in
 $\mathfrak T_1$   and in $\mathfrak T_2$, respectively.

As is well known, a crucial notion of quantum theory and of
quantum information is the concept of {\em entanglement\/}.
Consider a composite quantum system $S= S_1 + \ldots + S_n$.
According to the quantum theoretic formalism, the {\em reduced
state function\/} determines for any state $\rho$ of $S$ the {\em
reduced state\/} $Red^{i_1,\ldots,i_m}(\rho)$ of any subsystem
$S_{i_1}+ \ldots + S_{i_m}$ (where $1\le i_1 \le n, \ldots, 1\le
i_m \le n $.) A characteristic case that arises in
entanglement-phenomena is the following: while $\rho$ (the state
of the global system) is pure (a maximal information), the reduced
state $Red^{i_1,\ldots,i_m}(\rho)$ is generally a mixture (a
non-maximal information). Hence our information about the {\em
whole\/} cannot be reconstructed as a function of our pieces of
information about the {\em parts\/}.

In the second Part of this article we will see how these
characteristic {\em holistic\/} features of the quantum theoretic
formalism will play an important role in the development of the
epistemic semantics.

\begin{definition} {\em ($n$-partite entangled quregister)}
\label{de:npartito}\nl A quregister $\ket{\psi}$ of $\mathcal
H^{(n)}$ is called an {\em $n$-partite entangled\/}
 iff all reduced states
  $Red^1(\ket{\psi}),\ldots,Red^n(\ket{\psi})$ are proper mixtures.
\end{definition}

As a consequence an  $n$-partite entangled quregister cannot be
represented as
 a tensor product of the reduced states of its parts.

When all reduced states
$Red^1(\ket{\psi}),\ldots,Red^n(\ket{\psi})$ are the qumix
$\frac{1}{2}\Id$ (which represents a perfect
 ambiguous information) one says that $\ket{\psi}$ is {\em
 maximally entangled\/}.

\begin{definition} {\em (Entangled quregister with respect to some parts)}
\label{de:partentangl}\nl A quregister $\ket{\psi}$ of $\mathcal
H^{(n)}$ is called {\em entangled\/}
 with respect to its parts labelled by the
indices $i_1, \ldots,i_h$ (with $1 \le i_1,\ldots,i_h \le n$) iff
 the  reduced states $Red^{i_1}(\ket{\psi}),\ldots,
    Red^{i_h}(\ket{\psi})$ are proper mixtures.

\end{definition}

Since the notion of reduced state is independent of the choice of
a particular basis, it turns out that the status  of {\em
$n$-partite entangled quregisters\/},
 {\em maximally entangled quregisters\/}  and {\em entangled quregisters with
 respect to some parts\/} is invariant under changes of
 truth-perspective.

\begin{example}\nl

\begin{itemize}
\item The quregister
$$ \ket{\psi} = \frac{1}{\sqrt{2}}(\ket{0,0,0} + \ket{1,1,1}) $$
is a 3-partite maximally entangled quregister of $\mathcal
H^{(3)}$;
\item the quregister
$$ \ket{\psi} = \frac{1}{\sqrt{2}}(\ket{0,0,0} + \ket{1,1,0}) $$
is an entangled quregister  of $\mathcal H^{(3)}$  with
respect to its first and second part.
\end{itemize}
\end{example}


\section{Epistemic situations and epistemic structures}
Any logical analysis of epistemic phenomena naturally refers to a
set of agents (say,  Alice, Bob, ... ), possibly evolving in time.
Let $T = (\mathfrak t_1,\ldots, \mathfrak t_n)$ be a sequence of
times (which can be thought of as ``short'' time-intervals) and
let $Ag$ be a finite set of epistemic agents, described as
functions of the times in $T$. For any $\mathfrak a \in Ag$ and
any $\mathfrak t$ of $T$, we  write $\mathfrak a(\mathfrak t)=
\mathfrak {a_t}$.
  Each $\mathfrak a_\mathfrak t$ is associated with a
characteristic {\em epistemic situation\/}, which consists of the
following elements:
\begin{enumerate}
\item[1.] a truth-perspective $\mathfrak {T_{a_t}}$,
    representing the truth-conception of $\mathfrak a$ at time
    $\mathfrak t$.
\item[2.] A set $EpD_{\mathfrak {a_t}}$ of qumixes,
    representing the information that is virtually accessible
    to $\mathfrak {a_t}$ (a kind of {\em virtual memory}).
\item[3.] Two epistemic maps $\mathbf U_{\mathfrak {a_t}}$ and
    $\mathbf K_{\mathfrak {a_t}}$, that permit us to transform
    any qumix living in a space $\mathcal H^{(n)}$ into a
    qumix living in the same space. From an intuitive point of
    view, $\mathbf U_{\mathfrak {a_t}} \rho$ is to be
    interpreted as: $\mathfrak {a_t}$ understands $\rho$ (or,
   $\mathfrak {a_t}$ has information about $\rho$);  while
   $\mathbf K_{\mathfrak {a_t}} \rho$ is to be interpreted as:
    $\mathfrak {a_t}$ knows $\rho$. \end{enumerate}

    \begin{definition}{\em (Epistemic situation)}
    \label{de:epsit}\nl An {\em epistemic situation\/} for an
    agent $\mathfrak {a_t}$ is a system
    $$EpSit_{\mathfrak {a_t}} = (\mathfrak {T_{a_t}},\,
    EpD_{\mathfrak {a_t}},\,\mathbf U_{\mathfrak {a_t}},\,
    \mathbf K_{\mathfrak {a_t}}),  $$
    where:
    \begin{enumerate}
\item [1.] $\mathfrak {T_{a_t}}$ is a truth-perspective,
    representing the truth-conception of $\mathfrak {a_t}$.
\item[2.] $EpD_{\mathfrak {a_t}}$ is a set of qumixes,
    representing the virtual memory of $\mathfrak {a_t}$. We
    indicate by  $EpD^{(n)}_{\mathfrak {a_t}}$  the set
    $EpD_{\mathfrak {a_t}} \cap \mathfrak D(\mathcal
    H^{(n)})$.

\item [3.] $\mathbf U_{\mathfrak {a_t}}$ is a map that assigns
    to any $n\geq 1$ a map, called {\em (logical)
    understanding operation\/}:
$$\mathbf U^{(n)}_{\frak {a_t}}: \mathcal B(\mathcal
    H^{(n)}) \mapsto \mathcal B(\mathcal H^{(n)}),$$ where
    $\mathcal B(\mathcal H^{(n)})$ is the set of all bounded
    operators of $\mathcal H^{(n)}$. The following conditions
    are required:
    \begin{enumerate}
\item[3.1.]$\rho \in \mathfrak D(\mathcal H^{(n)}) \,\,
    \Longrightarrow \,\, \mathbf U^{(n)}_{\frak {a_t}}\rho
    \in \mathfrak D(\mathcal H^{(n)})$.
\item[3.2.] $\rho \notin EpD^{(n)}_{\mathfrak
    {a_t}}\,\,\Longrightarrow \,\,  \mathbf U^{(n)}_{\frak
    {a_t}}\rho = \overline{\rho_0}$ (where
    $\overline{\rho_0}$ is a fixed element of
  $\mathfrak D(\mathcal H^{(n)})$.

\end{enumerate}

\item [4.] $\mathbf K_{\mathfrak {a_t}}$ is a map that assigns
    to any $n\geq 1$ a map, called {\em (logical) knowledge
    operation\/}:
$$\mathbf K^{(n)}_{\frak {a_t}}: \mathcal B(\mathcal
    H^{(n)}) \mapsto \mathcal B(\mathcal H^{(n)}).$$
     The following conditions are required:
    \begin{enumerate}
\item[4.1.]$\rho \in \mathfrak D(\mathcal H^{(n)}) \,\,
    \Longrightarrow \,\, \mathbf K^{(n)}_{\frak {a_t}}\rho
    \in \mathfrak D(\mathcal H^{(n)})$.
\item[4.2.] $\rho \notin EpD^{(n)}_{\mathfrak
    {a_t}}\,\,\Longrightarrow \,\,  \mathbf K^{(n)}_{\frak
    {a_t}}\rho = \overline{\rho_0}$ (where
    $\overline{\rho_0}$ is a fixed element of
  $\mathfrak D(\mathcal H^{(n)})$.
\item [4.3.] $\mathbf K^{(n)}_\mathfrak {a_t} \rho
    \preceq_{\mathfrak {T_{a_t}}} \rho$, for any $\rho \in
    EpD^{(n)}_{\mathfrak {T_{a_t}}}$ (where
    $\preceq_\mathfrak {T_{a_t}}$ is the preorder relation
    defined by Def. \ref{de:preordine}).
\item [4.4.] $\mathbf K^{(n)}_\mathfrak {a_t} \rho
    \preceq_{\mathfrak {T_{a_t}}} \mathbf
    U^{(n)}_\mathfrak {a_t}\rho$, for any $\rho \in
    EpD^{(n)}_{\mathfrak a_t}$.

\end{enumerate}

    \end{enumerate}

    \end{definition}

For the sake of simplicity, we will generally write $\mathbf
U_\mathfrak {a_t}\rho$ and $\mathbf K_\mathfrak {a_t}\rho$,
instead of $\mathbf U^{(n)}_\mathfrak {a_t}\rho$ and $\mathbf
K^{(n)}_\mathfrak {a_t}\rho$.

According to Def. \ref{de:epsit},  whenever an information
  $\rho$ does not belong to the epistemic domain of
  $\mathfrak {a_t}$, then both $\mathbf U_{\mathfrak {a_t}}\rho$
  and $\mathbf K_{\mathfrak {a_t}}\rho$
  collapse into a fixed   element
  (which may be identified, for instance, with the maximally
  uncertain
  information  $\frac{1}{2}\Id^{(n)}$ or with the
   $\mathfrak {T_{a_t}}$-Falsity $^{\mathfrak {T_{a_t}}}P_0^{(n)}$ of the space
  $\mathcal H^{(n)}$ where $\rho$ lives). At the same time, whenever
  $\rho$  belongs to the epistemic domain of $\mathfrak {a_t}$,
   it seems
reasonable to assume that the probability-values of $\rho$ and
$\mathbf K_{\mathfrak {a_t}}\rho$ are correlated: the probability
of the quantum information asserting that ``$\rho$ is known by
$\mathfrak {a_t}$'' should always be less than or equal to the
probability of $\rho$ (with respect to the truth-perspective of
$\mathfrak {a_t}$) (condition 4.3.). Hence, in particular, we
have:
$$ {\tt p}_{\mathfrak {T_{a_t}}}(\mathbf K_{\mathfrak {a_t}}\rho) =1
 \,\,\,\Rightarrow
\,\,\,\tt p_{\mathfrak {T_{a_t}}}(\rho)=1. $$ But generally, not the
other way around! In other words, pieces of quantum information
that are known  are true (with respect to the truth-perspective of
the agent in question). Also condition 4.4. appears quite natural:
knowing implies understanding.

A  knowledge operation  $\mathbf K_{\mathfrak {a_t}}$  is called
{\em non-trivial\/}
   iff for at least one qumix $\rho$,
${\tt p}_{\mathfrak {T_{a_t}}}(\mathbf K_{\mathfrak {a_t}}\rho) <
{\tt p}_{\mathfrak {T_{a_t}}}(\rho)$. Notice that  knowledge
operations do not generally preserve pure states \cite{BDCGLS}.

For any agent $\mathfrak {a_t}$ whose  epistemic situation is
$(\mathfrak {T_{a_t}},\,
    EpD_{\mathfrak {a_t}},\,\mathbf U_{\mathfrak {a_t}},\,
    \mathbf K_{\mathfrak {a_t}})$, two special sets play an
    important intuitive role. The first set represents a kind of
    {\em active memory\/} of $\mathfrak {a_t}$, and can be defined as
    follows:
    $$ActMem(\mathfrak {a_t}) : = \parg{\rho \in EpD_{\mathfrak{a_t}}:
    {\tt p}_{\mathfrak {T_{a_t}}}(\mathbf U_{\mathfrak {a_t}}\rho) =1  }.  $$
    While the epistemic domain of $\mathfrak {a_t}$ represents
    the virtual memory of $\mathfrak {a_t}$,  $ActMem(\mathfrak
    {a_t})$ can be regarded as the set  containing all pieces of
    information that are actually
    understood by  agent  $\mathfrak {a}$  at time $\mathfrak
    {t}$. Another important set, representing the {\em actual knowledge\/}  of
    $\mathfrak a$ at time $\mathfrak t$, is
    defined as follows:
    $$ActKnowl(\mathfrak {a_t}) : = \parg{\rho \in EpD_{\mathfrak{a_t}}:
    {\tt p}_{\mathfrak {T_{a_t}}}(\mathbf K_{\mathfrak {a_t}}\rho) =1  }.  $$

    By definition of {\em epistemic situation\/} one immediately
    obtains:
    $$ActKnowl(\mathfrak {a_t}) \subseteq ActMem(\mathfrak {a_t})
    \subseteq EpD(\mathfrak {a_t}).  $$

    Using the concepts defined above, we can  now introduce the
notion of {\em epistemic quantum computational structure\/} (which
will play an important role in the development of the epistemic
semantics).

  \begin{definition} {\em (Epistemic quantum computational
  structure)}\label{de:epstruct} \nl
  An {\em epistemic   quantum computational   structure\/} is a system

 $$\mathcal S  = (T,\,Ag,\, \mathbf {EpSit})$$
 where:
 \begin{enumerate}
 \item [1.] $T$ is a time-sequence $(\mathfrak t_1,\ldots,
     \mathfrak t_n)$.
 \item [2.] $Ag$ is a finite set of epistemic agents
     $\mathfrak a$ represented as  functions of the times
     $\frak t$ in $T$.
 \item [3.] $\mathbf {EpSit}$ is a map that assigns to any
     agent $\mathfrak {a}$  at time $\mathfrak t$ an epistemic
     situation $EpSit_{\mathfrak {a_t}} = (\mathfrak
     {T_{a_t}},\, EpD_{\mathfrak {a_t}},\,\mathbf I_{\mathfrak
    {a_t}},\, \mathbf K_{\mathfrak {a_t}}) $.

  \end{enumerate}
  \end{definition}

It may happen that, at any time, all agents of an epistemic
quantum computational structure $\mathcal S$ share one and the
same truth-perspective. In other words, for any agents $\mathfrak
a$, $\mathfrak b$ and for any times $\mathfrak t_i$, $\mathfrak
t_j$: $\mathfrak {T_{a_{t_i}}} = \mathfrak {T_{b_{t_j}}}$.
 In such a case we will say that $\mathcal S$ is {\em
(epistemically) harmonic}.

 It is interesting to isolate some characteristic properties that
 may be satisfied by the agents of an epistemic quantum computational structure.

 \begin{definition} \label{de:capacities}\nl
 Let $\mathcal S  = (T,\,Ag,\, \mathbf {EpSit})$ be an
 epistemic quantum computational structure and let $\mathfrak a$ be an agent of
 $\mathcal S$.
 \begin{itemize}
 \item $\mathfrak a$ has a {\em sound epistemic capacity\/}
     iff for any time $\mathfrak t$, the qumixes $^{\mathfrak
     {T_{a_t}}}P_1^{(1)}$ and  $^{\mathfrak
     {T_{a_t}}}P_0^{(1)}$ belong to the epistemic domain of
     $\mathfrak {a_t}$. Furthermore, $ \mathbf K_{\mathfrak
     {a_t}}\,^{\mathfrak {T_{a_t}}}P_1^{(1)}\,\, = \,\,
     ^{\mathfrak {T_{a_t}}}P_1^{(1)}$ and $ \mathbf
     K_{\mathfrak {a_t}}\,^{\mathfrak {T_{a_t}}}P_0^{(1)}\,\,
     = \,\, ^{\mathfrak {T_{a_t}}}P_0^{(1)}$. In other words,
     at any time,  agent $\mathfrak a$ has access to the
truth-values  of his/her truth-perspective, assigning to them
     the ``right'' probability-values.
  \item $\mathfrak a$ has a {\em perfect epistemic capacity\/}
      iff for any time $\mathfrak t$ and any qumix $\rho$
      belonging to the epistemic domain of $\mathfrak {a_t}$,
$\mathbf K_{\mathfrak {a_t}}\rho = \,\, \rho$. Hence, at any
     time
$\mathfrak a$ assigns the ``right'' probability-values to all
      pieces of information that belong to his/her epistemic
      domain.
  \item $\mathfrak a$ has a {\em maximal epistemic capacity\/}
      iff,  at any time $\mathfrak t$, $\mathfrak a$ has a
      perfect epistemic capacity and his/her  epistemic domain
      coincides with the set $\mathfrak D$
  of all possible qumixes.
\end{itemize}
 \end{definition}

 Notice that a maximal epistemic capacity does not imply {\em
 omniscience\/} (i.e. the capacity of {\em deciding\/} any piece
 of information). For,  in quantum computational logics
 the {\em excluded-middle principle}
 $$\forall \rho \in \mathfrak D(\mathcal H^{(n)}):\,\,\,
 \text{either}\,\,\, {\tt p}_\mathfrak T(\rho) = 1 \,\,\,
 \text{or}\,\,\,
 {\tt p}_\mathfrak T(^\mathfrak D{\tt NOT}_\mathfrak T^{(n)}\rho) = 1 $$
 is, generally, violated.

 When all agents of an epistemic quantum computational structure
 $\mathcal S$ have a sound (perfect, maximal) capacity, we will say
 that $\mathcal S$ is {\em sound\/} ({\em perfect\/}, {\em
 maximal\/}).

 In many  concrete epistemic situations agents use to {\em interact\/}. In order to describe
 such phenomenon from an abstract point of view, we introduce the
 notion of  {\em epistemic quantum computational structure with
 interacting agents\/}.

 \begin{definition} {\em (Epistemic quantum computational structure with
                    interacting agents)} \label{de:interact}  \nl
 An  {\em epistemic quantum computational structure with
 interacting agents} is a system
 $$\mathcal S  = (T,\,Ag,\, \mathbf {EpSit},\, Int),$$
 where:
 \begin{enumerate}
 \item [1.] $(T,\,Ag,\, \mathbf {EpSit})$  is an epistemic
     quantum computational structure;
 \item [2.] $Int$ is a map that associates to any time
     $\mathfrak t \in T$ a set of pairs $\left(\mathfrak a_t,
     \mathfrak b_t\right)$ (where $\mathfrak a,\, \mathfrak b
 \in Ag$). The intuitive interpretation of $\left(\mathfrak
 a_\mathfrak t, \mathfrak b_\mathfrak t\right) \in
 Int(\mathfrak t)$ is: the agents $\mathfrak a$ and $\mathfrak
 b$ interact at time $\mathfrak t$;
 \item [3.] $\left(\mathfrak a_\mathfrak t, \mathfrak
     b_\mathfrak t\right) \in Int(\mathfrak t) \Rightarrow$\nl
     $ \exists \mathfrak t' \geq \mathfrak t \exists \rho
     [(\rho \in ActMem(\mathfrak {a_t}) \,\,\text{and}\,\,
     \rho \in ActMem(\mathfrak {b_{t'}}))$\nl $\text{or}\,\,
     (\rho \in ActMem(\mathfrak {b_t}) \,\,\text{and}\,\, \rho
     \in ActMem(\mathfrak {a_{t'}}))]$.\nl In other words, as
     a consequence of the interaction,
there is at least one piece of information $\rho$ such that at
        time $\mathfrak t$  agent $\mathfrak a$ certainly
        understands $\rho$, while at a later time $\mathfrak
        t'$  agent $\mathfrak b$ certainly understands $\rho$;
        or viceversa.

 \end{enumerate}

 \end{definition}

 What can be said about the characteristic mathematical properties
 of epistemic operations? Is it possible to represent the knowledge operations
  $\mathbf K^{(n)}_{\mathfrak {a_t}}$ occurring in an
  epistemic quantum computational structure as special cases of qumix gates?
  This question has a negative answer.
 One can prove that non-trivial  knowledge
operations cannot be represented by unitary quantum operations
\cite{BDCGLS}.

 At the same time, some interesting knowledge
operations can be represented by the more general notion of {\em
quantum channel\/} (which represents a special case of the concept
of {\em quantum operation\/}\footnote{See for instance \cite{CDP}
and \cite {Hong}.}).

\begin{definition} {\em (Quantum channel)}\label{quantochann}\nl
A {\em quantum channel\/} on $\mathcal H^{(n)}$ is a linear map
$\mathcal E$ from  $\mathcal B(\mathcal H^{(n)})$ to $\mathcal
B(\mathcal H^{(n)})$  that satisfies the following properties:
\begin{itemize}\item for any $A \in \mathcal B(\mathcal H^{(n)})$,
    ${\tt Tr}(\mathcal E(A)) = {\tt Tr}(A)$;
\item $\mathcal E$ is completely positive.
\end{itemize}

\end{definition}

From the definition one immediately obtains that any quantum
channel maps qumixes into qumixes.

A useful characterization of quantum channels is stated by  {\em
Kraus first representation theorem\/} \cite{K}.

\begin{theorem}
A map
 $$\mathcal E: \mathcal B(\mathcal H^{(n)}) \mapsto
\mathcal B(\mathcal H^{(n)})$$
 is a quantum channel on  $\mathcal H^{(n)}$ iff for some set $I$ of indices there exists a
 set
 $\parg{E_{i}}_{i \in I}$ of elements of $\mathcal B(\mathcal H^{(n)})$
 satisfying the following conditions:
 \begin{enumerate}
 \item $\sum_i E_i^\dagger E_i = \Id^{(n)}$;
 \item $\forall A \in  \mathcal B(\mathcal H^{(n)}): \mathcal
     E(A) = \sum_i E_iAE_i^\dagger   $.

 \end{enumerate}

\end{theorem}

  Of course, qumix gates $^\mathfrak DG^{(n)}$ are special cases
   of quantum channels,
 for which
  $\parg{E_{i}}_{i \in I}= \parg{G^{(n)}}$.

  One can prove that there exist uncountably many quantum channels
  that are non-trivial  knowledge operations of the
 space $\mathcal H^{(n)}$ with respect to any truth-perspective \cite{BDCGLS}.

 An interesting example  of a quantum channel that gives rise to a
a knowledge operation is the {\em depolarizing
 channel\/}. Let us refer to the space $\mathcal H^{(1)}$
and let $p\in[0,1]$. Consider the following system of operators:
$$E_0=\frac{\sqrt{4-3 p}}{2}\,\Id^{(1)};\,
E_1=\frac{\sqrt{p}}{2}{\tt X};\, E_2=\frac{\sqrt{p}}{2}{\tt Y};\,
E_3=\frac{\sqrt{p}}{2}{\tt Z},$$ (where ${\tt X}$, ${\tt Y}$,
${\tt Z}$ are the three Pauli-matrices). Define $^p\mathcal
{D}^{(1)}_\mathfrak T$ as follows for any $\rho \in \mathfrak
D(\C^2)$:
$$\, ^p\mathcal {D}^{(1)}_\mathfrak T \rho=\sum_{i=0}^3 \mathfrak T E_i\mathfrak T^\dagger\,\rho\,\mathfrak T E_i^{\dagger}\mathfrak T^\dagger.$$

It turns out that  $^p\mathcal{D}^{(1)}_\mathfrak T$ is a quantum
channel,  called {\em depolarizing channel\/}. Notice that for any
truth-perspective
 $\mathfrak T$, $^p\mathcal {D}^{(1)}_\mathfrak T=\,^p\mathcal
 {D}^{(1)}_\Id$.

  The channel $^p\mathcal {D}^{(1)}_{\tt I}$ gives
 rise to a corresponding knowledge operation
  $^p\mathbf {KD}^{(1)}_{\mathfrak {a_t}}$ for an agent $\mathfrak {a_t}$
  (who
  is supposed to belong to an epistemic quantum computational
  structure $\mathcal S$).

\begin{definition} \emph{(A depolarizing knowledge operation
$^p\mathbf {KD}^{(1)}_{\mathfrak {a_t}}$)} \label{def:dep}\nl
Define $^p\mathbf {KD}^{(1)}_{\mathfrak {a_t}}$ as follows:
\begin{enumerate}
\item $EpD_{\mathfrak{a_t}} \subseteq \parg{\rho\in \mathfrak
    D(\mathcal H^{(1)}):\Prob_{\mathfrak {T_{a_t}}}(\rho)\geq
    \frac{1}{2}}$.
\item $\rho \in EpD_{\mathfrak {a_t}}\;\Rightarrow \;
    \,\,^p\mathbf {KD}_{\mathfrak{a_t}} ^{(1)}\rho =
    \,^p\mathcal {D}^{(1)}_{{\tt I}} \rho$.
\end{enumerate}
\end{definition}
Consider now $^1\mathbf {KD}_{\mathfrak {a_t}}^{(1)}$ and suppose
that the structure $\mathcal S$ satisfies the condition:
  $$\rho \notin EpD_{\mathfrak {a_t}}
   \,\,\Rightarrow \,\, ^1\mathbf {KD}_{\mathfrak {a_t}}^{(1)}\rho =
   \frac{1}{2}{\tt I}^{(1)}.$$ We obtain:
for any $\rho\in \mathfrak D(\mathcal H^{(1)})$, $^1\mathbf
{KD}_{\mathfrak {a_t}}^{(1)}\rho =\,\, ^1\mathcal {D}^{(1)}_{{\tt
I}} \rho\, = \, \frac{1}{2}{\tt I}^{(1)}$.\nl In other words,
$^1\mathbf {KD}_{\mathfrak {a_t}}^{(1)}$ seems to behave like a
``fuzzification-procedure'', that transforms any (certain or
uncertain) knowledge into a kind of maximally unsharp piece of
information.

Other examples  of quantum channels representing knowledge
operations that  give rise to interesting  physical
interpretations have been investigated in \cite{LStoappear}.

Unlike qumix gates, knowledge operations  are not generally
reversible. One can guess that the intrinsic irreversibility of
the {\em act of knowing\/} is somehow connected with a loss of
information due to
 the interaction with an environment.

\section{Memorizing and retrieving information via teleportation}
In epistemic processes that concern both human and artificial
intelligence it is customary to distinguish an {\em internal\/}
from an {\em external memory\/}. In the framework of our approach,
 the internal memory $IntMem_{\mathfrak {a_t}}$ of an agent
$\mathfrak a$ (say, Alice) at time $\mathfrak t$ can be naturally
associated with the set $ActMem(\mathfrak {a_t})$. Hence, a piece
of information $\rho$ will belong to the internal memory of
$\mathfrak {a_t}$ iff ${\tt p}_{\mathfrak {T_{a_t}}}(\mathbf
U_{\mathfrak {a_t}} \rho) = 1$. This means that at time $\mathfrak
t$  Alice has a kind of ``aware understanding'' of the information
$\rho$. At the same time, the external memory $ExtMem_{\mathfrak
{a_t}}$, can be identified with a convenient subset  of the
epistemic domain of $\mathfrak {a_t}$. Owing to the concrete
limitations of the  internal memory, the possibility of
``depositing elsewhere'' (in an external memory) some pieces of
information turns out to be very useful for  Alice. Of course, at
a later time, Alice  should be able to retrieve her ``forgotten''
information, storing it again in her internal memory.

We will now try to model examples  of this kind in the framework
of our abstract quantum computational approach. We will refer to a
very simple physical situation. At any time $\mathfrak t$ (of a
given time-sequence) the external memory $ExtMem_{\mathfrak
{a_t}}$ of  Alice is supposed to be physically realized by a
two-particle system $S_1 + S_2$, while the internal memory
$IntMem_{\mathfrak {a_t}}$ is realized by a single particle $S_3$.
For any time $\mathfrak t$, the global system $S(\mathfrak t)=
(S_1+S_2+S_3)(\mathfrak t)$ will represent  Alice's {\em physical
memory-system\/}. For the sake of simplicity, we suppose that  the
state of $S(\mathfrak t)$, indicated by $\ket{\Psi^S(\mathfrak
t)}$, is pure. Accordingly, $\ket{\Psi^S(\mathfrak t)}$ will
determine the states of the subsystems, which will be, generally,
mixtures. We write: \nl $\rho^{(S_1+S_2)}(\mathfrak t)=
Red^{(1,2)}(\ket{\Psi^S(\mathfrak t)}); $\nl
$\rho^{(S_i)}(\mathfrak t)= Red^{(i)}(\ket{\Psi^S(\mathfrak t)}) $
(where $1\le i \le 3$).

On this basis, we can put:
\begin{itemize}
\item $IntMem_{\mathfrak {a_t}}= \parg{\rho^{(S_3)}(\mathfrak
    t)}$;
\item $ExtMem_{\mathfrak {a_t}}=
\parg{\rho^{(S_1)}(\mathfrak
    t), \rho^{(S_2)}(\mathfrak t), \rho^{(S_1+S_2)}(\mathfrak
    t)}$.

\end{itemize}

Now we want to describe a process of ``memorizing and retrieving
information'', by using a quantum teleportation phenomenon.
Physically, this process corresponds to the time-evolution of the
global memory-system $S$ (in a given time-interval). Since during
this process  Alice's internal and external memories shall
interact, we can imagine that  Alice's  external memory is
associated with an agent $\mathbf b$ (say, Bob), who can
communicate with Alice via a classical channel (as happens in the
standard teleportation-cases). Accordingly,  our abstract
description will naturally make use of epistemic quantum
computational structures with interacting agents (Def.
\ref{de:interact}). For the sake of simplicity, we will refer to
harmonic structures, where all agents have,  at any time, the
canonical truth-perspective ${\tt I}$.
\begin{center} {\bf At time ${\mathfrak t_1}$} \end{center}
We suppose that at the initial time $\mathfrak t_1$ the global
memory-state is the following:
$$\ket{\Psi^S(\mathfrak t_1)}= \frac{1}{\sqrt{2}}(\ket{0,0}+ \ket{1,1})\,\otimes\,
(a_0\ket{0} + a_1\ket{1}). $$ Hence, the state of the external
memory is the entangled {\em Bell-state\/}, while the state of the
internal memory is a qubit. According to our convention, we
obtain:\nl  $IntMem_{\mathfrak {a_t}}=
\parg{\rho^{(S_3)}(\mathfrak
    t)}$, where
    $\rho^{(S_3)}(\mathfrak
    t)= P_{a_0\ket{0} + a_1\ket{1}}$;\nl
$ExtMem_{\mathfrak {a_t}}=
\parg{\rho^{(S_1)}(\mathfrak
    t), \rho^{(S_2)}(\mathfrak t), \rho^{(S_1+S_2)}(\mathfrak
    t)}$, where \nl
$\rho^{(S_1)}(\mathfrak
    t)= \frac{1}{2}{\tt I}^{(1)}$;\,\,
$\rho^{(S_2)}(\mathfrak
    t)= \frac{1}{2}{\tt I}^{(1)}$;\nl
$\rho^{(S_1+S_2)}(\mathfrak
    t)= P_{\frac{1}{\sqrt{2}}(\ket{0,0}+ \ket{1,1})}$.

    \begin{center} {\bf At time ${\mathfrak t_2}$} \end{center}
    In order to ``forget'' the information $a_0\ket{0} + a_1
    \ket{1}$ (stored by her internal memory) Alice
    acts on her global memory, by applying the gate
    ${\tt SWAP}_{(1,3)}^{(3)}$, which exchanges the states of the first
    and of the third subsystem of $S$.
    As a consequence, we obtain:
    $$\ket{\Psi^S(\mathfrak t_2)}=
    {\tt SWAP}_{(1,3)}^{(3)} \ket{\Psi^S(\mathfrak t_1)}=
    (a_0\ket{0} + a_1\ket{1})\otimes \frac{1}{\sqrt{2}}(\ket{0,0}+\ket{1,1}). $$
    Alice's internal memory is now  changed. We have:
    $$\rho^{(S_3)}(\mathfrak t_2) = \frac{1}{2}{\tt I}^{(1)},$$ which represents
     a maximally fuzzy information. Roughly, we might
     say that at time $\mathfrak t_2$  Alice has ``cleared
     out'' her internal memory. At the same time, we have that
     $\rho^{(S_1)}(\mathfrak t_2) = P_{a_0\ket{0} + a_1\ket{1}}$
     belongs to  Alice's external memory. The operation of
     memorizing the  information $a_0\ket{0} + a_1\ket{1}$ in the
     external memory is  now completed.
     Interestingly enough, the
     entanglement correlation between $S^{(3)}(\mathfrak t_2)$ and
     $S^{(2)}(\mathfrak t_2)$ guarantees  to Alice the
     possibility  of interacting with her external
     memory.
     It turns out  that the transformation
     $\rho^{(S_3)}(\mathfrak t_1) \, \mapsto
      \, \rho^{(S_3)}(\mathfrak t_2)$ is  described by the
        depolarizing  knowledge operation
         (considered in the previous
        Section), which transforms any $\rho$ of $\mathcal H^{(1)}$  into
      $\frac{1}{2}{\tt I}^{(1)}$.

     Notice that the state of the global system
     $\ket{\Psi^S(\mathfrak t_2)}= (a_0\ket{0} + a_1\ket{1})\otimes
     \frac{1}{\sqrt{2}}(\ket{0,0}+\ket{1,1})$ corresponds to the
     initial state of the standard teleportation-situation, where
     Bob (who has physical access to the system $S_1+S_2$)
     tries to send the qubit $a_0\ket{0} +
     a_1\ket{1}$ to the
      ``far'' Alice (who has access to $S_3$),  by using the
     entanglement-correlation between $S_2$ and $S_3$. We can now
     proceed, by applying the  steps that are currently used in a
     teleportation-process.

     \begin{center} {\bf At time ${\mathfrak t_3}$} \end{center}
Bob  applies the gate ${\tt XOR}^{(1,1)}$ to the external
memory-state. As a consequence, we obtain:\nl
$\ket{\Psi^S}_{t_3}=\left[{\tt XOR}^{(1,1)}\otimes {\tt I}^{(1)}
\right] \ket{\Psi^S}_{t_2}=$\nl
$\frac{1}{\sqrt{2}}\left(a_0\ket{0}\otimes(\ket{0,0} +
\ket{1,1}\right))+
\frac{1}{\sqrt{2}}\left(a_1\ket{1}\otimes(\ket{1,0} +
\ket{0,1}\right)).
 $

It is worth-while noticing that {\em theoretically\/}   Bob is
acting on the whole system $S$, while {\em materially\/} he is
only acting  on the subsystem $S_1 + S_2$ that is accessible to
him.

\begin{center} {\bf At time $\mathfrak t_4$} \end{center}\nl
 Bob applies the gate Hadamard to the system $S_1$ (whose
state is to be teleported into the internal memory). Hence, we
obtain: \nl $\ket{\Psi^S}_{t_4}=\left[\sqrt{{\tt I}}^{(1)}\otimes
{\tt I}^{(1)} \otimes {\tt I}^{(1)}\right] \ket{\Psi^S}_{t_3}=$\nl
$ \frac{1}{2}[(\ket{0,0}\otimes (a_0\ket{0} + a_1\ket{1})) \otimes
(\ket{0,1}\otimes (a_0\ket{1} + a_1\ket{0}))+$\nl
$(\ket{1,0}\otimes (a_0\ket{0} - a_1\ket{1})) + (\ket{1,1}\otimes
(a_0\ket{1} - a_1\ket{0}))].$

\begin{center} {\bf At time $\mathfrak t_5$} \end{center}\nl
 Bob performs a measurement on the external memory, obtaining
as a result one of the  following possible registers:
  $\ket{0,0},\,\ket{0,1},\,\ket{1,0},\, \ket{1,1}$.
  As a consequence,  the state of the global system is transformed,
  by {\em collapse of the wave-function\/};
   and such transformation is mathematically described by a
   (generally irreversible) quantum operation.

   Let $P^{(2)}_{\ket{x,y}}$ represent the projection-operator
   over the closed subspace determined by the register
   $\ket{x,y}$. We obtain four possible states for the global
   memory-system:
\begin{enumerate}
\item [1.] $\ket{\Psi_{00}^S(\mathfrak t_5)}=
    2[P^{(2)}_{\ket{0,0}}\otimes {\tt I}^{(1)}]
    \ket{\Psi^S(\mathfrak t_4)}= \ket{0,0}\otimes (a_0\ket{0}
    + a_1\ket{1})$;
\item [2.] $\ket{\Psi_{01}^S(\mathfrak t_5)}=
    2[P^{(2)}_{\ket{0,1}}\otimes {\tt I}^{(1)}]
    \ket{\Psi^S(\mathfrak t_4)}= \ket{0,1}\otimes (a_0\ket{1}
    + a_1\ket{0})$;
\item [3.] $\ket{\Psi_{10}^S(\mathfrak t_5)}=
    2[P^{(2)}_{\ket{1,0}}\otimes {\tt I}^{(1)}]
    \ket{\Psi^S(\mathfrak t_4)}= \ket{1,0}\otimes (a_0\ket{0}
    -a_1\ket{1})$;
\item [4.] $\ket{\Psi_{11}^S(\mathfrak t_5)}=
    2[P^{(2)}_{\ket{1,1}}\otimes {\tt I}^{(1)}]
    \ket{\Psi^S(\mathfrak t_4)}= \ket{1,1}\otimes (a_0\ket{1}
    -a_1\ket{0})$.

\end{enumerate}

By quantum non-locality,  Bob's action on the external
    memory  has determined an instantaneous transformation of  the
    state $\rho^{(S_3)}(\mathfrak t_4)$ of the internal memory,
     which will have now one
    of the four possible forms:
    $$a_0\ket{0}+ a_1\ket{1};\,\, a_1\ket{0}+ a_0\ket{1};\,\,
    a_0\ket{0} - a_1\ket{1};\,\, a_0\ket{1}- a_1\ket{0}.   $$
Alice's internal memory is no longer fuzzy, since it is storing
again a qubit. However, only in the first case this qubit
coincides with the original $a_0\ket{0}
    + a_1\ket{1}$ that Alice  had stored in her internal
    memory (at the initial time). In spite of this, Alice
    has the possibility of retrieving her original information,
    through the application of a convenient gate. We have:\nl
  $a_0\ket{0} + a_1\ket{1} = {\tt I}^{(1)}(a_0\ket{0} +
  a_1\ket{1})
  = {\tt NOT}^{(1)}(a_1\ket{0} + a_0\ket{1})=
  {\tt Z}(a_0\ket{0} - a_1\ket{1})
  = {\tt NOT}^{(1)}{\tt Z}(a_1\ket{0} - a_0\ket{1})$
(where ${\tt Z}$ is the third Pauli-matrix).

In this situation, Bob  can give an ``order'' to  Alice, by using
a classical communication channel. The order will be:
\begin{itemize}
\item ``apply ${\tt I}^{(1)}\,$!'' (i.e. ``don't do
    anything!''), in the first case.
\item ``apply ${\tt NOT}^{(1)}\,$!'', in the second case.
\item ``apply ${\tt Z}^{(1)}\,$!'', in the third case.
\item ``apply ${\tt NOT}^{(1)}{\tt Z}\,$!'' in the fourth
    case.

\end{itemize}
\begin{center} {\bf At time $\mathfrak t_6$} \end{center}\nl
 Alice follows Bob's order and retrieves her original
information.

 Notice that the transformation
$\rho^{(S_3)}(\mathfrak t_1) \,\, \mapsto
\,\,\rho^{(S_3)}(\mathfrak t_6)$ (from the initial to the final
state of the internal memory) is mathematically described by the
identity operator. Transformations of this kind (which concern
reduced states and are obtained by neglecting the interaction with
an environment) generally determine a loss of information;
consequently they are described by irreversible quantum
operations. Interestingly enough, this is not the case in the
situation we have considered here, where the
entanglement-correlation between the internal and the external
memory, associated with a classical communication, allows Alice to
retrieve {\em exactly\/} her initial information.

\end{document}